\documentclass[openacc]{rstransa}

\usepackage{booktabs}
\usepackage{graphicx,subfigure}
\usepackage{dcolumn}
\usepackage{float}
\usepackage{bm}

\begin{document}

\title{Evaporation Induced Rayleigh-Taylor Instabilities in Polymer Solutions}

\author{
E. J. Mossige$^{1,*}$, V. Chandran Suja$^{1,*}$, M. Islamov$^{2}$, S. F. Wheeler$^{1}$ and G. G. Fuller $^{1}$}

\address{$^{1}$ Department of Chemical Engineering, Stanford University, CA - 94305, USA\\
$^{2}$ Department of Chemical Engineering,Columbia University, New York, NY 10027, USA\\
$^{*}$ Equal Contribution}

\subject{Fluid mechanics}


\corres{Gerald G. Fuller \\
\email{ggf@stanford.edu}}

\begin{abstract}
Understanding the mechanics of detrimental convective instabilities in drying polymer solutions is crucial in many applications such as the production of film coatings. It is well known that solvent evaporation in polymer solutions can lead to Rayleigh-B{\'e}nard or Marangoni-type instabilities. Here we reveal another mechanism, namely that evaporation can cause the interface to display Rayleigh-Taylor instabilities due to the build-up of a dense layer at the air-liquid interface. We study experimentally the onset time ($t_p$) of the instability as a function of the macroscopic properties of aqueous polymer solutions, which we tune by varying the polymer concentration ($c_0$), molecular weight and polymer type. In dilute solutions, $t_p$ shows two limiting behaviors depending on the polymer diffusivity. For high diffusivity polymers (low molecular weight), the pluming time scales as $c_0^{-2/3}$.  This result agrees with previous studies on gravitational instabilities in miscible systems where diffusion stabilizes the system. On the other hand, in low diffusivity polymers the pluming time scales as $c_0^{-1}$. The stabilizing effect of an effective interfacial tension, similar to those in immiscible systems, explains this strong concentration dependence. Above a critical concentration, $\hat{c}$, viscosity delays the growth of the instability, allowing time for diffusion to act as the dominant stabilizing mechanism. This results in $t_p$ scaling as $(\nu/c_0)^{2/3}$. 


\end{abstract}


\begin{fmtext} 

\end{fmtext} 

\maketitle \noindent 
\section{Introduction}

In the production of polymer films, uniformity in thickness and concentration is paramount. There are numerous physical processes that interfere with this desired result such as skin formation\cite{komoda2014local,okuzono2006simple} and convective motion 
 such as thermal Rayleigh-B{\'e}nard \cite{berg1966natural} and Marangoni instabilities \cite{rodriguez2019evaporation, suja2018evaporation,shin2016benard,sakurai2002control,bormashenko2008mesoscopic}. These latter examples generate convective flow cells that redistribute polymers anisotropically within the film, potentially leading to inhomogeneities  \cite{bassou2008role}.  

We report here a different mechanism for producing nonuniformities in polymer solutions caused by evaporation. Instead of inducing convective flow cells, we find that the action of solvent evaporation can induce an unstable concentration profile that triggers a gravity-driven Rayleigh-Taylor (RT) instability \cite{lord1900investigation,taylor1950instability}. We utilize aqueous polymer solutions that we pour into open cavities that are exposed to the ambient air. As water evaporates from these water-based solutions, an increasingly dense polymer-rich layer forms at the surface. Above a critical density difference with the underlying bulk, this layer becomes unstable to a gravitational mixing process. 

The RT instability in miscible systems is observed in other applications. Evaporation induced RT instability was first reported by the Colinet group using ethanol-water mixtures \cite{dehaeck2009evaporating}. As alcohol escapes from these "evaporating cocktails", a water-rich surface layer develops, creating an unstable stratified system. Other examples include the geotaxis of microorganisms \cite{plesset1974bioconvection,dunstan2018evaporation}, the dissolution of minerals in freshwater \cite{julien2018dissolutionPRF} and mixing with seawater \cite{ramaprabhu2004experimental}, granular media flows \cite{vinningland2007granular}, and absorption of $CO_2$ in brine solutions for carbon capture and storage applications \cite{thomas2018convective}. Mixing in the accidental painting technique \cite{zenit2014} is an example of how artists exploit density differences between pigmented layers of paint to create interesting patterns. Reviews on the RT instability in miscible environments are given by Andrews and Dalziel \cite{andrews2010small} and by Boffetta and Mazzino \cite{boffetta2017incompressible}.


In this work we examine the mechanism of evaporation induced RT instabilities in polymer solutions. In addition to being important to numerous technical applications, polymer systems offer the advantage of systematically studying the influence of viscosity on evaporation induced instabilities. Furthermore, as presented below, they allow one to examine the transition between diffusion controlled and diffusion free behavior. In addition, with the correct choice of polymer, one can avoid the occurrence of elastic film formation. 

Most studies concerning the RT instability have treated stratified layers of \emph{immiscible} fluids. 
When the upper fluid is more dense than the lower one, the interface between the two fluids becomes unstable with a fastest growing wavelength dictated by the capillary length \cite{kull1991theory}.  The interfacial tension, $\sigma$, separating the fluids gives rise to a capillary force that stabilizes against short wavelengths, while wavelengths above a certain threshold can grow unbounded, allowing irregularities to amplify. 

RT instabilities in miscible systems have the distinction that interfacial tension forces, in the traditional sense, are absent. 
Density differences that lead to RT instabilities produce two stabilizing effects. On the one hand, diffusion of individual components will oppose concentration nonuniformities. Alternatively, concentration gradients can also have a stabilizing influence by giving rise to surface stresses that mimic surface tensions in \emph{immiscible} systems \cite{lacaze2010transient,truzzolillo2017off,Truzzolillo_2017,joseph1993fluid}. 
For most miscible systems, including polymer solutions, the effective interfacial tension $\Gamma_e$ can be calculated using the Korteweg formula, $\Gamma_e$=$\zeta$/${\Delta c^2}{d}$ \cite{korteweg1901forme,truzzolillo2016prx}, where $\Delta c$ is the concentration contrast, $\zeta$ is a system dependent fitting parameter and $d$ is the interface thickness. Accounting for the transient stresses is important in order to accurately describe the dynamics of miscible interfaces, such as that of miscible drops rising or descending in another miscible fluid \cite{chen2001miscible, chen2002miscible}. Recently, Gowda et al. \cite{gowda2019effective} exploited transient stresses in a flow-focusing microchannel to stabilize the interface between a colloidal fibre suspension and its own solvent. Transient stresses can also delay or even prevent hydrodynamic instabilities in miscible liquids \cite{Truzzolillo_2017,lyubimova2019rayleigh}. For example, sharp concentration gradients can prevent fingering instabilities between miscible liquids of different viscosities \cite{truzzolillo2014prl}. 

This paper explores the relative importance of these two stabilizing mechanisms against evaporation induced RT instabilities in aqueous polymer solutions. Polymer solutions offer the advantage of systematic variation of the relative importance of diffusion, viscosity and surface stresses through a variation of polymer type, molecular weight and concentration. The observable in our experiment is the onset time at which visible instabilities in concentration appear. A successful scaling analysis is able to correlate this observable against the controlled variables in the experiments that displays the transition between diffusion controlled and diffusion free behavior. The scaling analysis can be generalized to other binary systems including colloidal suspensions and saline solutions.

\section{Methods}
\begin{figure}
    \centering
    \includegraphics[width = \linewidth]{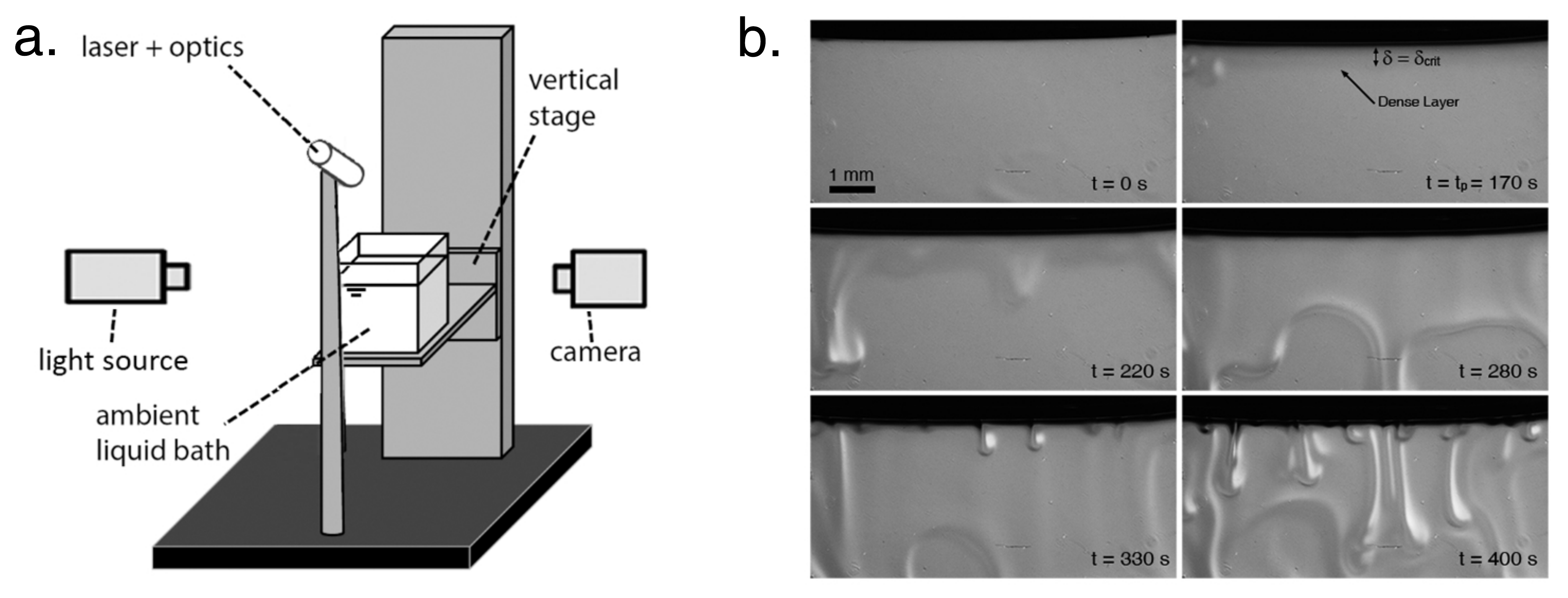}
    \caption{{\bf a.} Schematic of the experimental setup. A backlight imaging system enables visualization of the RT instability and consists of a platform to hold the test chamber containing a polymeric solution, a light source that produces collimated light that illuminates the sample, and a camera that captures images of the mixing process. {\bf b.} Sequence of images (time stamps on the bottom right) showing the development of the evaporation induced Rayleigh-Taylor instability in an aqueous polymer solution containing $20 wt\%$ $40$kDa Dextran. The first plumes become visible around $t=170 s$ and marks the onset time for pluming of the sample, $t_p$. $\delta$ denotes the thickness of the growing dense layer at the air-liquid interface.}
    \label{fig:plumeseq}
\end{figure} 
In order to learn about the evaporation induced RT instability, we utilize two oft-studied aqueous systems containing Dextran and (poly)ethylene glycol (PEG) polymers. To tune the physical properties of these water-based solutions, we use polymer molecular weights of 6, 40 and 500 kDa for Dextran and molecular weights of 10 and 35 kDa for PEG. We also vary the initial polymer concentration, $c_0$, between $0.1$ and $20\;wt\%$. These two polymers are ideal for two reasons. First, skin formation does not occur since the interfacial storage modulus , $G_s^{'}$, is negligible ( see details in  Supplementary Materials). Second, these polymer solutions behave as simple viscous liquids in our experiments and do not exhibit viscoelastic or shear thinning effects. This is because the characteristic shear rates in our experiments are small, resulting in very small Deborah numbers and corresponding Newtonian behavior. For details on the rheological response of the polymer solutions see Supplementary Materials.


In a typical experiment, we fill these polymer solutions into an open test cell made from microscope slides that were glued together.  The cell is sufficiently large in the spanwise direction (Width $\times$ Depth = 9.9 $\times$ 5.5 $mm^2$) to ensure that influences from side walls can be neglected. In addition, the liquid column height of $h = 3.2\;mm$ results in small values of the Marangoni and thermal Rayleigh numbers (see Appendix), thus ensuring thermal Rayleigh-B{\'e}nard-Marangoni instabilities have only a minimal influence on our measurements. We revisit the implications of the channel dimensions on the measurement accuracy in the discussion section. Finally, the solutions are exposed to air at room temperature ($22 \pm 2^oC$, $50 \pm 7\%$ RH). With these stable experimental conditions, water evaporates at an even rate, yielding an evaporation velocity at the air-liquid interface of $\Dot{h} = 5 \pm 1.57 \times 10^{-7}\; m/s $. $\dot{h}$ was determined by analyzing videos of the evaporating solutions and was found to be independent of the tested polymer types and concentrations. 


A backlight imaging system serves to visualize the RT instability, see Fig.\ref{fig:plumeseq} (a). The setup consists of a central platform (Model: ULM-TILT, Newport, CA, USA) to hold the test chamber, a $2x$ telecentric lens (Model: 63074, Edmund Optics, NJ, USA) and camera (Model: GPF 125C IRF, Allied Vision Technologies, PA, USA) to capture images, and a light source (Model: 21AC fiber optic illuminator, Edmund Optics, NJ, USA) attached to a telecentric backlight collimator (Model: 62760, Edmund Optics, NJ, USA) to illuminate the samples. The platform is fixed on a linear stage (Model: ULM-TILT, Newport, CA, USA) to align the chamber in front of the camera. A detailed description of the setup can be found in Ref. \cite{walls2018shape}.

\section{Results}
\subsection{Visualizations of the instability}
Fig.\ref{fig:plumeseq} (b) shows a sequence of images from a typical experiment utilizing a polymer solution containing $20\;wt\%$ 40 kDa Dextran.  The time stamp $t=0$ in the top left image denotes the state immediately after filling. As water evaporates, a dense layer grows until it reaches a critical thickness, $\delta_{crit}$. At this point, the layer becomes RT-unstable, which is apparent from the descending plume-like structures within the solution. These plumes, as well as the dense layer, can be seen in the tile with the time stamp $t=t_p=170\;s$. The first plumes emanate from a sheet-like structure ($t = 220\;s$ and $t = 280\;s$), followed by a second regime where the number of plumes more than doubles. As long as  evaporation is present, this second regime is sustained at long times ($t = 330\;s$ and $t = 400\;s$). 



\subsection{Onset time for the instability}

In this study, the observable we report is the onset time of the instability (see the tile with the time stamp $t=t_p=170\;s$ in Fig.\ref{fig:plumeseq} (b)) with reference to the filling time at  $t=0$. This observable is measured as a function of molecular weight and concentration for two different polymer materials.

\begin{figure}[]
    \centering
    \includegraphics[width =  \linewidth]{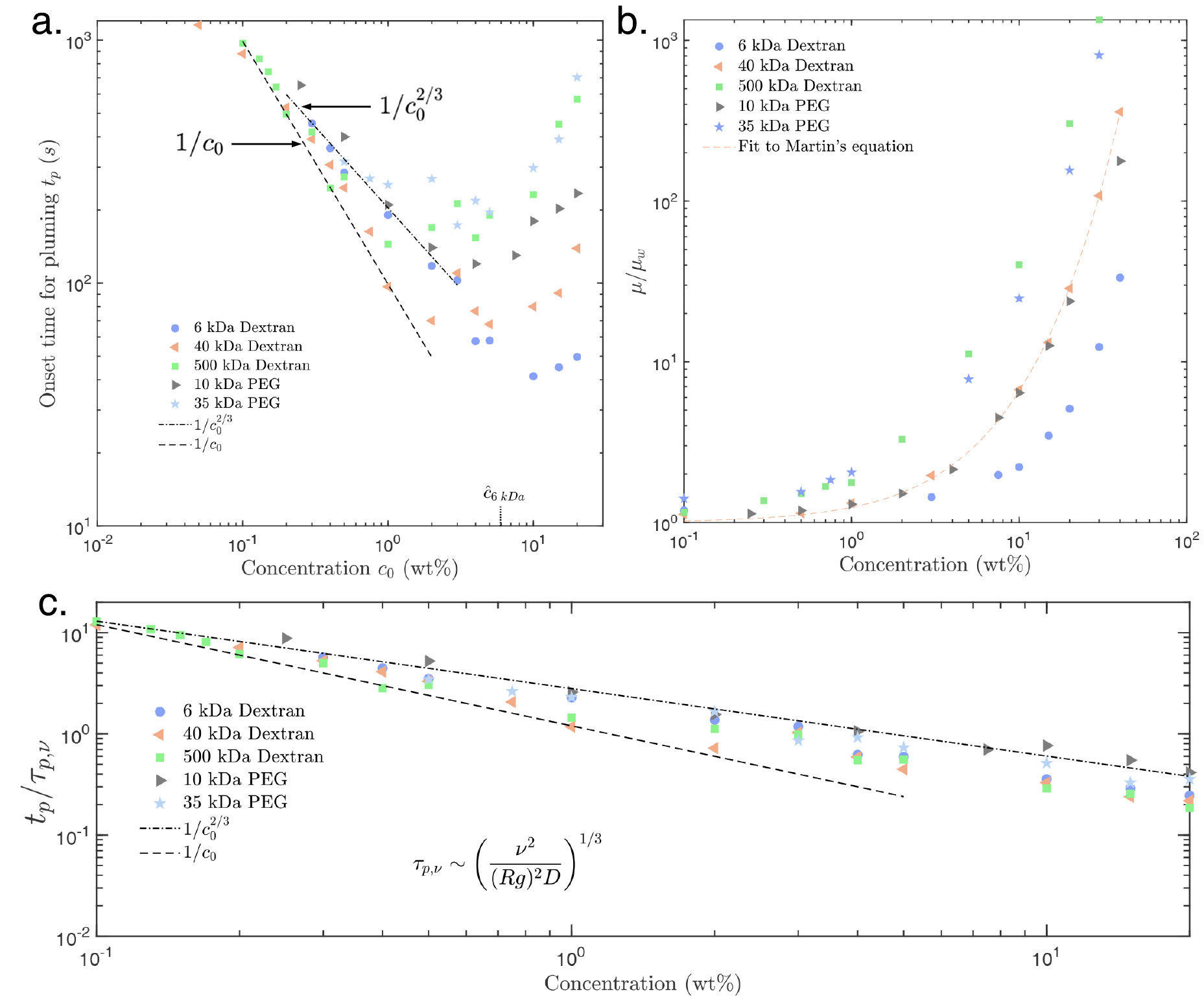}
    \caption{{\bf a.} Experimentally obtained critical time for the onset of the Rayleigh-Taylor instability, $t_p$, for different polymers as a function of their concentration, $c_0$. Each data point represents the mean value of three independent measurements. The lines indicate the two limiting behaviors observed in the experiments, namely diffusion free ($t_p \propto c_0^{-1}$) and diffusion-controlled ($t_p \propto c_0^{-2/3}$). $\hat{c}_{6\;kDa}$ indicates the concentration that leads to a minimum $t_p$ for $6\;kDa$ Dextran. {\bf b.} Dynamic viscosities for the solutions as functions of polymer concentration.  The viscosities increase exponentially with respect to concentration and follow the prediction by Martin \cite{Martin1942, weissberg1951viscosity}. The viscosities are made dimensionless by the viscosity of the solvent (water). {\bf c.} The onset times of convection, $t_p$, when scaled with a viscous time scale $\tau_{p,\nu} \propto \nu^{2/3}$, where $\nu$ is the kinematic viscosity of the solution. The decay of the scaled onset times are bound between $-1$ and $-2/3$ (indicated by the broken lines).}
    \label{fig:time2plume}
\end{figure}

Fig. \ref{fig:time2plume} (a) shows the dependency of the onset time $t_p$ on the initial polymer concentration $c_0$ for two different polymer materials, PEG and Dextran. Each data point represents the mean value of at least three independent measurements with a typical standard deviation of $20\;\%$. In the log-log plot in Fig. \ref{fig:time2plume} (a), $t_p$ first decreases monotonically until a critical concentration, after which it begins to increase. The rate of diminution depends on the polymer molecular weight. Further, the concentration where the onset time reaches a minimum is inversely correlated to the polymer molecular weight. For example, $t_p$ for the largest polymer, Dextran 500 kDa, reaches a minimal value at a concentration of 1$\%$, while $t_p$ for the smallest molecule, Dextran 6 kDa, has a minimal value at a concentration of 10$\%$.


The presence of a minimum in $t_p$ in Fig.\ref{fig:time2plume} (a) suggests that there is competition between two physical mechanisms, where one mechanism dominates below a critical concentration, $c_0$ < $\hat{c}$, while another mechanism dominates above $\hat{c}$. It is possible to obtain $\hat{c}$ from bulk viscosity measurements (Fig. \ref{fig:time2plume} (b)) and we present such a procedure in the Apprendix. In Fig. \ref{fig:time2plume} (a), we indicate the value of $\hat{c} = \hat{c}_{6kDa}=9.2\%$ for the smallest polymer molecule, namely Dextran 6 kDa. It is apparent from Fig.\ref{fig:time2plume} (a) that below $\hat{c}$, the pluming times have a power-law dependence on concentration. The two dashed lines in the figure indicate an upper and lower bound to that power-law dependence that will be described below. We also describe the deviation of $t_p$ from these power laws for concentrations exceeding $\hat{c}$. 



\subsubsection{Diffusion free regime: Dilute solutions containing long chained polymers}


In miscible systems containing large molecules, diffusion is too slow compared to the convective motion of the interface to act as an effective stabilizer. Instead, transient stresses arising from sharp concentration gradients stabilize the interface through an effective interfacial tension\cite{truzzolillo2017off}. The industrial applications of stable miscible interfaces are many. For example, effective interfacial tensions are utilized to align nanofibrils in co-flowing microchannels to create sustainable materials \cite{gowda2019effective}. In our experiments, sharp concentration gradients between the polymer-rich surface layer and the the bulk can also have a stabilizing effect. In diluted solutions containing long chained polymers, diffusion does not smear out concentration differences on the time scale of the instability, which results in the interface remaining well defined. Conveniently, in this regime, we can use theory developed for immiscible fluids to analyze the RT instability.  


In the immiscible RT instability, the competition of forces is between gravity, which de-stabilizes the interface, and surface tension, which is a stabilizing mechanism. The ratio of these competing forces is the Bond number, $Bo$=$\Delta \rho g l^2$/$\sigma$, where $\sigma$ is the surface tension between the immiscible fluids, $l$ the wavelength of the instability, $g$ the acceleration due to gravity, and $\Delta \rho$ the density difference between the interfacial layer and the bulk. In miscible systems without diffusion, effective interfacial tensions, $\Gamma_e$, take the role of $\sigma$ yielding $Bo$=$\Delta \rho g l^2$/$\Gamma_e$. In open systems, as studied here, the density difference is time dependent, $\Delta \rho$, yielding a time-dependent Bond number, $Bo(t)$=$\Delta \rho(t) g l^2$/$\Gamma_e$. When $Bo(t)$ exceeds a critical value, the system is said to be unstable. Assuming negligible diffusion and that the changes in $\Gamma_e$ both as a function of time and polymer concentration can be neglected, the onset time for the instability ($t_p$) is then solely governed by the time it takes to build up a fixed critical density difference ($\Delta \rho_c$).

\noindent By definition, the far field (initial) concentration $c_0$ and density $\rho_0$ are:
\begin{align}
    c_0 = \frac{m_p}{m_p+m_w}, \\
    \rho_0 = \frac{m_p+m_w}{m_p/\rho_p+m_w/\rho_w} = \rho_w \frac{ \eta + 1}{\eta(\rho_w/\rho_p)+1}. \label{eq.densitydef}
\end{align}
Here, $m_p$ and $\rho_p$ are the mass and density of the polymer, respectively, and $m_w$ and $\rho_w$ are the corresponding quantities for water and $\eta = c_0/(1-c_0)$. Since evaporation is depleting water from the system at a rate of $\Dot{h}$, we can cast Eq.\ref{eq.densitydef} as function of $\Dot{h}$ and surface area $A$ as follows:
\begin{equation}
    \rho(t) = \frac{m_p+m_w - \Dot{h}At}{m_p/\rho_p+(m_w -\Dot{h}At) /\rho_w}.
\end{equation}
Note that we have neglected polymer diffusion when writing the above expression. Using the definition of $\eta$ and rearranging yields:
\begin{equation}
    \rho(t) = \rho_w \frac{(\eta + 1)m_w- \Dot{h}At}{(\eta (\rho_w/\rho_p)+ 1) m_w -\Dot{h}At}.
\end{equation}

\noindent The onset time for pluming is identically equal to the time at which $\rho(t) - \rho_0 = \Delta \rho_c $:
\begin{equation}
    \Delta \rho_c = \rho_w \frac{(\eta + 1)m_w- \Dot{h}At_c}{(\eta (\rho_w/\rho_p)+ 1) m_w -\Dot{h}At_c} - \rho.
\end{equation}

\noindent Utilizing the definition of $\rho$ and noting $\Dot{h}A/m_w = \kappa$, we obtain the following expression for $t_p$, 
\begin{equation}
  t_p = \frac{\Delta \rho_c}{\rho_w \kappa}  \frac{\eta (\rho_w/\rho_p)+ 1}{\Delta \rho_c/\rho_w -1 + (\eta +1)/(\eta(\rho_w/\rho_p)+1)}.
\end{equation}
Assuming $\Delta \rho_c/\rho_w \ll 1$ (justified as our interfacial tensions are also quite small), we can further simplify the above expression and use the definition of $\eta$ to obtain, 

\begin{equation}
     t_p = \frac{\Delta \rho_c}{\rho_w \kappa} \frac{\rho_p}{\rho_p+\rho_w}\frac{1-c_0}{c_0}\left( \frac{\rho_w}{\rho_p}\frac{c_0}{1 -c_0} +1 \right)^2.
\end{equation}
Expanding the above expression in powers of $c_0$ and denoting $\frac{\rho_w}{\rho_p} = \Grave{\rho}$, we obtain 
\begin{equation}
    t_p = \frac{1}{\kappa} \frac{\Delta \rho_c \rho_p}{\rho_w(\rho_p+\rho_w)} \left[ \frac{1}{c_0} + (2\Grave{\rho} - 1 ) + \Grave{\rho}^2 c_0  + \mathcal{O}(c_0^2)  \right].
\end{equation}
We identify the prefactor as a modified Atwood number $At = \Delta \rho_c/(\rho_w +\rho_w \Grave{\rho})$, yielding:

\begin{equation}
    t_p = \frac{At}{\kappa} \left[ \frac{1}{c_0} + (2\Grave{\rho} - 1 ) + \Grave{\rho}^2 c_0 + \mathcal{O}(c_0^2) \right].
    \label{eq:tau_pe}
\end{equation}
For dilute solutions  $t_p \propto 1/c_0$ at leading order, which is in alignment with the experimental data for $c_0$ < $1\%$ for $500$ kDa Dextran (Fig.\ref{fig:time2plume} (a)).

 
\subsubsection{Diffusion limited regime}
\begin{figure}
    \centering
    \includegraphics[width =  \linewidth]{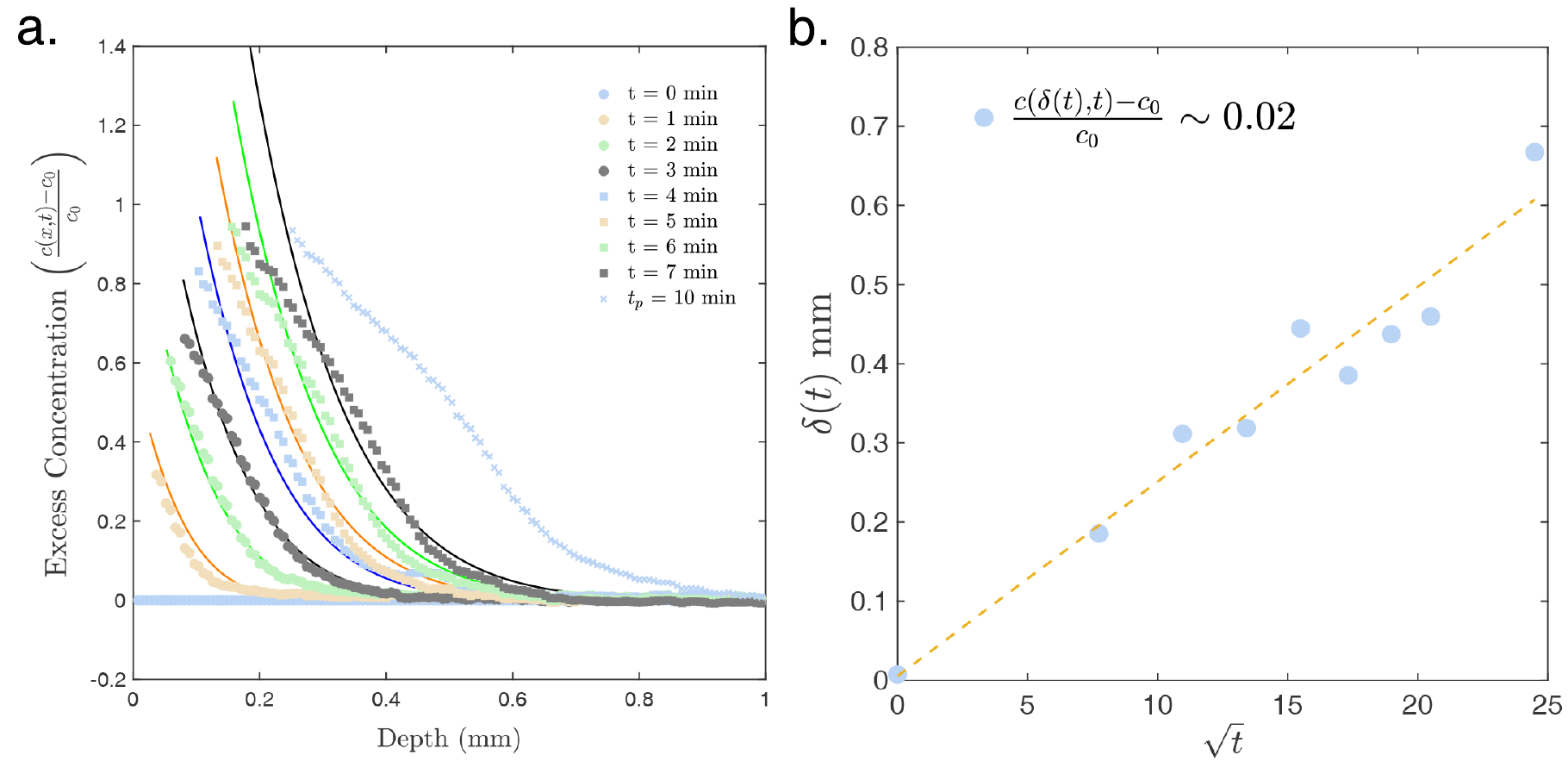}
    \caption{{\bf a.} Experimentally obtained diffusion profiles in the dense layer developing near the air-liquid interface prior to the instability for $20\%$ $500\; kDa$ Dextran in water (solid circles). Here the depth is measured from the position of the free interface at time $t=0$. The solid lines are the theoretical predictions for diffusion profiles (Eq.\ref{eq.DiffusionEqCompleteSolution}) utilizing  diffusion coefficient ($\mathcal{D} = 1.1\times 10^{-10}\;m^2/s$) and evaporation velocities ($\Dot{h} = 4\times 10^{-7}\; m/s $) measured $\textit{in situ}$ (see Supplementary Materials). {\bf b.} The experimentally obtained thickness, $\delta$, of the dense layer as a function of $\sqrt{t}$. The proportionality constant relating $\delta$ and $\sqrt{Dt}$, is numerically obtained as $\sim \;2.4$ for the reported evaporation velocity.}
    \label{fig:concentrationProfiles}
\end{figure}

However, the rest of the dataset in Fig.\ref{fig:time2plume} (a) does not align with the prediction of Eq.\ref{eq:tau_pe}. The deviation from the $1/c_0$ decay is especially prominent for small polymers (low molecular weight) and above $\hat{c}$ for all tested polymers. To explain this behavior, it is helpful to note that a prominent feature that precedes the onset of the instability is an evolving dense layer (Fig.\ref{fig:plumeseq}; $t=170s$). Utilizing Beer-Lambert's law \cite{swinehart1962beer}, the pixel intensities can be converted to polymer concentration within this layer. The filled symbols in Fig.\ref{fig:concentrationProfiles} (a) show the resulting profiles of one such measurement ($20\;wt\%$ 500 kDa Dextran solution) at different times $t$. By plotting the the thickness of the dense layer as a function of $t$ (Fig.\ref{fig:concentrationProfiles}b), we observe that $\delta \propto \sqrt{t}$, indicating a diffusive process. 

To confirm this diffusive behavior and to exclude other effects on the evolution of the concentration profile, we compare the experimental observation to a theoretical prediction utilizing a 1-D diffusion equation \cite{okuzono2006simple}. Solving this equation on a reference frame convecting with the air-liquid interface moving at a constant speed, $\dot{h}$, yields:
\begin{multline}\label{eq.DiffusionEqCompleteSolution}
    \frac{c(x, t) - c_0}{c_0} = \left[\left(2 \alpha^2 t- \frac{\dot{h} x}{2\mathcal{D}} +\frac{1}{2} \right) \text{erfc}(\beta) e^{-\frac{\dot{h}x}{\mathcal{D}}} - \frac{1}{2}\text{erfc}(\gamma) + 2 \alpha \sqrt{\frac{t}{\pi}} e^{-\beta^2}e^{-\frac{\dot{h}x}{\mathcal{D}}} \right], 
\end{multline} where 
\[\alpha = \frac{\Dot{h}}{2\sqrt{\mathcal{D}}}, \quad \beta = \frac{x}{2\sqrt{\mathcal{D}t}} - \alpha \sqrt{t}, \quad \gamma = \frac{x}{2\sqrt{\mathcal{D}t}} + \alpha \sqrt{t}.  \]
The lines in Fig.\ref{fig:concentrationProfiles}(a) are theoretical predictions and are calculated from Eq.\ref{eq.DiffusionEqCompleteSolution} using values of $\dot{h}$ and $\mathcal{D}$ of $4.4\times 10^{-7}\; m/s$ and $1.1\times 10^{-10}\; mm^2/s$, respectively. These values accurately capture the diffusive growth of the dense layer and were chosen to best fit the measured profiles. This value of $\dot{h}$ is within the experimental uncertainty of the measured value of $5\pm1.57\times 10^{-7}\; m/s$. Likewise the value of $\mathcal{D}$ also accurately fits the growth of the dense layer (Fig.\ref{fig:concentrationProfiles} (b)). As time approaches the pluming time $t_p =10 $ min, the dense layer shows irregularities, and as expected, Eq.\ref{eq.DiffusionEqCompleteSolution}  no longer reproduces the measured profile. As more water evaporates, $\delta$ eventually achieves a critical value of $\delta_{crit} \propto\sqrt{\mathcal{D} t_p}$ causing the instability. This diffusive growth is in contrast to the case of dilute solutions of $500$ kDa Dextran polymers, where diffusion can be neglected.

The RT instability in diffusive systems can be triggered only if the convective motion of descending plumes exceeds the rate at which diffusion dampens out the density differences with the underlying bulk liquid. In other words, the system is unstable when the ratio of the diffusive to the convective time scale exceeds a critical value, known as the Rayleigh number for species diffusion \cite{blair1969onset,mahler1970stability}:
\begin{equation} \label{eq.RaSpecies}
    Ra_{s} =\frac{g\delta^3R\Delta c}{\nu \mathcal{D}} \;\textgreater \; Ra_{crit}.
\end{equation}  

\noindent Here, $\nu$ is the kinematic viscosity, $R = \frac{1}{\rho} \frac{d \rho}{d c}$ is the solutal expansion coefficient, and $\Delta c=c(x=0,t_p) - c_0$ is the concentration difference between the air-liquid interface ($x=0$) and the bulk at time $t_p$. The critical thickness, $\delta_{crit} \propto\sqrt{\mathcal{D} t_p}$, corresponds to a critical Rayleigh number, $Ra_{crit}$. Following previous works on gravitational instabilities in miscible systems \cite{blair1969onset,mahler1970stability,kurowski1995gravitational}, we substitute $\delta_{crit}$ in Eq. \ref{eq.RaSpecies} to obtain a time scale for the onset of the instability, 
\begin{equation} \label{eq.CharacteristicPlumingTime}
    t_{p} \propto \left( \frac{\nu}{gR\Delta c\mathcal{D}^{1/2}}\right)^{2/3} = \frac{\tau_{p,\nu}}{(\Delta c)^{2/3}},
\end{equation}
where $\tau_{p,\nu}$=$\nu^{2/3}$/($(R g)^{2/3} \mathcal{D}^{1/3})$ is a characteristic viscous time scale.

Eq. \ref{eq.CharacteristicPlumingTime} reveals two mechanisms where the pluming time depends on concentration. The diffusive growth of the dense layer leads to the power law decrease in pluming time as $\Delta c^{-2/3}$. Concentration also enters through the strong dependence of viscosity on this variable. The measured viscosities (see Fig.\ref{fig:time2plume}(b)) increase exponentially with concentration and fit well to Martin's equation \cite{Martin1942, weissberg1951viscosity}. The diffusivity $\mathcal{D}$ also depends on concentration, but is known that it has a much weaker dependence, $\partial \nu/\partial c \gg \partial\mathcal{D}/\partial c$ \cite{vergara1999mutual,walls2018shape, kim2000prediction}. For this reason, we set $\mathcal{D}$ in Eq.\ref{eq.CharacteristicPlumingTime} equal to a constant without loss of accuracy (the value of $\mathcal{D}$ is given in Table \ref{suptab:propWater}). The same argument is true for density (density changes by $< 10\%$).  Scaling $t_p$ by the viscous time $\tau_{p,\nu}$ substantially collapses the data (Fig.\ref{fig:time2plume}(c)). Based on this successful scaling, evidently viscosity is responsible for the increasing pluming time at higher concentration by retarding the growth rate of the instability.

The pluming times, when scaled with $\tau_{p,\nu}$, fall between the limits of $c_0^{-1}$ and $c_0^{-2/3}$. The transition from diffusion-free to diffusion controlled behavior is most readily apparent by considering the data collected for the $500$ kDa Dextran. Below  $1\;wt\%$, the pluming times descend inversely proportional to concentration, and then follow the power law $c_0^{-2/3}$. The data for the smallest molecules for each polymer, $6$ kDa Dextran and  $10$ kDa PEG, follow the diffusion controlled scalings for all the concentrations measured. The pluming times of $40$ kDa Dextran and $35$ kDa PEG display behavior that is intermediate between diffusion-free and diffusion controlled.

\section{Discussion}
The present paper focuses on the onset time of the instability. Another important parameter in the RT instability is the length scale of the instability, that is, the wavelength. Both diffusion ($\mathcal{D}$) and viscosity ($\nu$) modify the fastest growing wavenumber, $\lambda$, which goes as  $\lambda \, \sim \, (\mathcal{D}\nu/gR)^{1/3}$ as  explained by Kurowski and co-workers  \cite{kurowski1995gravitational}. As previously mentioned, in our system, $\mathcal{D}$ and $R$ are both insensitive to changes in concentration as compared to $\nu$ with respect to the same variable. As a result, the fastest growing wavelength becomes a function of viscosity alone, resulting in $\lambda \, \sim \, \nu^{1/3}$ in our system. 

The extent of the chamber along the optical path (Depth=5.5 $mm$) is larger than the wavelength $\lambda$. This is to ensure that influences from side walls can be neglected, as explained in the Methods section. An unfortunate consequence of the large depth as compared to $\lambda$ is that the observed plumes can exist in planes that are at different distances from the camera.  Therefore, $\lambda$ cannot be accessed directly from the videos. However, in line with the theoretical prediction, our results qualitatively show a strong correlation between the observed wavelength and viscosity (see Supplementary  Materials). 



A biological analogue to evaporating polymer solutions concerns non-motile bacteria suspended in saline solutions in open containers. As water evaporates, a dense and salt-enriched surface layer forms above the underlying solution. Bacteria within this layer fall with the descending plumes at several times their individual Stokes' velocities, effectively utilizing the RT instability as a mode of transportation. 

The Goldstein group were the ones to discover the mobility mechanism described above. Using physiological (dilute) saline concentrations ranging between 0.1 and 1 $\%$ they find that the onset times are "8 times longer for a 10 times more dilute suspension" \cite{dunstan2018evaporation}. This dependence of the onset time on concentration yields a power law exponent of $-0.9$, falling well within the limits of $-1$ and $-2/3$ developed in our paper.


The agreement between the onset times reported for saline solutions containing non-motile bacteria and our prediction should not come as a surprise as there are clear analogies. Firstly, the same mechanism induces the instability in both cases, namely the creation of a dense surface layer through solvent evaporation. Naturally, the salinity Rayleigh number for saline systems corresponds to the mass-based Rayleigh number in the present paper. In both cases, this parameter must exceed a critical value in order for the RT instability to occur. However, there are differences between saline solutions subject to evaporation and the present study. Most notably, a focus of our paper is to characterize the sensitivity of the instability with respect to changes in viscosity. By using polymer solutions, we were able to vary this parameter by three orders of magnitude. The viscosities of saline solutions are much less sensitive to changes in concentration. As a result, it is not possible to cover the same parameter space using saline solutions. 


Note that the instability reported here is distinct from the Marangoni instability or thermal Rayleigh-B{\'e}nard (RB) instabilities. In the case of the former instability the pluming scale is found to scale as, $t_{p} \propto \left( \nu /\Delta c\right)^2$ \cite{blair1969onset}, which is outside the range of decay we observe in our experiments; it is clear that Marangoni instabilities are not causing the pluming events. RB-instabilities are not present in our experiments for the following two reasons. First, the Lewis number is $\sim \mathcal{O}(10^3)$, suggesting gradients in temperature are dampened more quickly than gradients in density. Second, the maximal thermal Rayleigh number, $Ra_{max} = 680$, is below the critical value of $1000$ \cite{toussaint2008experimental} required to trigger the RB instability. The value of $Ra_{max}$ is based on a temperature differential of $0.5^oC$ between the air-liquid surface and the bulk and is typical in these water based systems, as reported previously \cite{spangenberg1961convective}. See Table \ref{suptab:propWater} in the Appendix for a lists of the material properties used in these calculations. 

\section{Conclusion}
We show that Rayleigh-Taylor (RT) instabilities can occur in polymer solutions subject to evaporation and in the absence of skin formation. For concentrated solutions, the onset time of the instability scales with viscosity $(\nu)$ and bulk concentration ($c_0$), yielding $t_p \propto (\nu/c_0)^{2/3}$. For dilute solutions, the onset time is bound between two limits: a diffusion limited regime where $t_p \propto 1/c_0^{2/3}$, and a diffusion free regime where $t_p \propto  1/c_0$. When the scaling analysis developed in this paper is applied non-motile bacteria in saline solutions undergoing RT instabilities, similar power-law dependencies are found. 


\section*{Appendix}
\subsection*{Calculation of the thermal Rayleigh number, Marangoni number and the Lewis number}

The maximal thermal Rayleigh number, $Ra_{max}$, is obtained using water without polymers (Table \ref{suptab:propWater}) in the test chamber (height $h=5\; mm$). With a surface to bulk temperature differential of $\Delta T = 0.5\;K$ \cite{spangenberg1961convective}, $Ra_{max}$ is:
\[ Ra_{max} = \frac{\beta \Delta T \rho g h^3}{\mu \alpha} = 680.\]

\noindent The Lewis number is
\[ L_e = \frac{\alpha}{\mathcal{D}} \sim 10^3. \]

\noindent The Marangoni number for the same conditions is, 
\[Ma = \frac{(-d\gamma/dT)\Delta T h}{\mu \alpha}  = 2622, \]

\begin{table}[!h]
\centering
\begin{tabular}{@{}lll@{}}
\toprule
Property                                & Units    & Value               \\ \midrule
Thermal expansion coefficient ($\beta$) & $1/K$    & $2.07\times10^{-4}$ \cite{berg1966natural} \\
Thermal diffusivity ($\alpha$)          & $m^2/s$  & $1.43\times10^{-7}$ \cite{berg1966natural}\\
Density ($\rho$)                        & $kg/m^3$ & $1000$              \\
Dynamic viscosity ($\mu$)               & $kg/m/s$ & $1\times10^{-3}$           \\
Diffusion coeff. of $500\;kDa$ Dextran ($\mathcal{D}$)        &  $m^2/s$ & $1.1\times 10^{-10}$\\
\bottomrule
\end{tabular}
\caption{Properties of water used for calculation of the thermal Rayleigh number and the Lewis number. Optical concentration profile measurements are used to obtain the diffusion coefficient for $500\;kDa$ Dextran.} \label{suptab:propWater}
\end{table}
\noindent 
\enlargethispage{20pt}

\subsection*{Method used to determine the concentration $\Hat{c}$}

The viscosity-concentration relation of the tested polymer solutions are described by Martin's equation \cite{Martin1942,weissberg1951viscosity}:

\begin{equation}
    \mu/\mu_w = 1 + c[\mu] e^{K_H [\mu]c },
    \label{eq.Martin}
\end{equation}
where $c$ is the weight based concentration, $[\mu]$ is the intrinsic viscosity, $K_H$ is the Huggins parameter and $\mu_w$ is the viscosity of the solvent (water in our case).

\begin{figure}[ht]
    \centering
    \includegraphics[width = 0.65\linewidth]{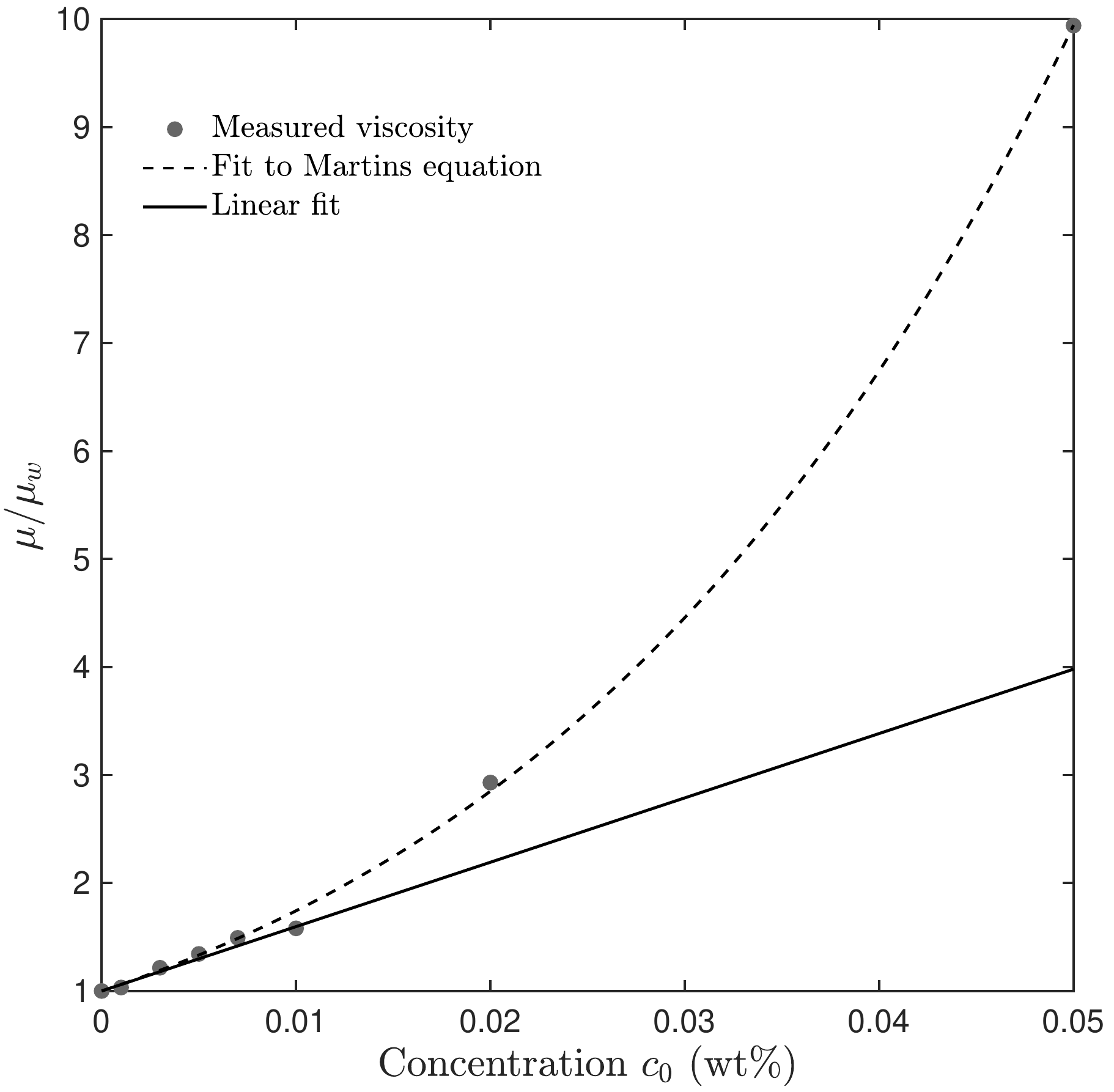}
    \caption{Plot showing measured values of the non-dimensional solvent viscosity $\mu/\mu_w$ of $500\;kDa$ Dextran solution (symbols) up to $5\;\%$ concentration. $\mu_w$ is the solvent (water) viscosity. The broken line is an exponential fit to the measured data utilizing Martin's equation. The solid line is a linear approximation, obtained by truncating the exponential after the leading order term (this term is a constant). The exponential and the linear fit deviate by $25\;\%$ at a concentration of $2.5\;\%$, which we identify as $\hat{c}$. }
    \label{supfig:FindingCHatfor500kDaDextran}
\end{figure}

The exponential in Martin's equation (Eqn. \ref{eq.Martin}) can be expanded in power series yielding $\mu/\mu_w$ = 1 + $c [\mu]$ + $\mathcal{O}(K_H c^2 [\mu]^2)$. Here, the linear portion is the Einstein-viscosity owing to Brownian motion of individual polymer molecules at infinite dilution. For concentrations above this dilute limit, long-range hydrodynamical or direct interactions between polymers cause the viscosity to increase non-linearly with concentration. We are interested in identifying the transition point between these two regimes in order to isolate viscous effects on the pluming events. In order to do so, we identify a critical concentration, $\hat{c}$, at which point the non-linear Martin-viscosity exceeds the linear Einstein-prediction by a certain threshold. We choose this threshold to be $25\;\%$ for our purpose. Fig. \ref{supfig:FindingCHatfor500kDaDextran} is a plot of the reduced viscosities $\mu/\mu_w$ of $500 kDa$ Dextran up to $5\;\%$ concentration. The symbols represent the measured viscosities, and the dotted line is a fit to Eqn. \ref{eq.Martin}, showing good agreement. The solid line is the linear Einstein-viscosity, which deviates from Martin's equation by $25\;\%$ at a concentration of $c_0$ = $\hat{c}$ = $2.5\;\%$. The numerical values of $\hat{c}$ for the other polymer solutions, namely $6$ and $40 kDa$ Dextran, and $10$ and $35 kDa$ PEG, are $9.2\;\%$, $4.6\;\%$, $5.3\;\%$ and $2.8\;\%$, respectively.  

\newpage
\dataccess{A separate document contains supporting material.}

\aucontribute{EJM and VCS contributed equally. EJM carried out experiments and led the experimental investigation, analyzed the data and drafted the manuscript. VCS conceived of the study, carried out experiments, analyzed the data, developed the theory and drafted the manuscript. MI and SFW both carried out experiments and  analyzed the data. GGF conceived of and designed the study and improved the manuscript. All authors read and approved the manuscript.}

\competing{The authors declare that they have no competing interests.}


\ack{We thank Daniel Walls for help with setting up the initial experiments and Sam Dehaeck for useful discussions.}



\bibliographystyle{vancouver}
\bibliography{References}

\end{document}